\documentclass[review,12p,authoryear,times]{elsarticle}%
\usepackage{natbib}
\usepackage{color}
\usepackage[latin1]{inputenc}
\usepackage{graphics}
\usepackage{booktabs}
\usepackage{amsmath}
\usepackage{amssymb}
\usepackage{overpic}
\usepackage{booktabs}
\journal{Icarus}
\begin{document}
\begin{frontmatter}
\title{A note on the survival of the sungrazing Comet C/2011 W3 (Lovejoy) within the Roche limit}
\author[igep]{B. Gundlach}
\ead{b.gundlach@tu-bs.de}
\author[igep]{J. Blum}
\author[igep]{Yu. V. Skorov}
\author[igep]{H. U. Keller}
\address[igep]{Institut für Geophysik und extraterrestrische Physik, Technische Universität Braunschweig, \\Mendelssohnstr. 3, D-38106 Braunschweig, Germany}
\begin{abstract}
In this work, a novel approach to explain the survival of sungrazing comets within the Roche limit is presented. It is shown that in the case of low tensile strength of the cometary nucleus, tidal splitting of the nucleus can be prevented by the reaction force caused by the sublimation of the icy constituents. The survival of Comet C/2011 W3 (Lovejoy) within the Roche limit of the Sun is, thus, the result of high tensile strength of the nucleus, or the result of the reaction force caused by the strong outgassing of the icy constituents near the Sun.
\end{abstract}
\end{frontmatter}

\setlength{\tabcolsep}{10pt}
\renewcommand{\arraystretch}{2}
\renewcommand{\topfraction}{1.0}
\renewcommand{\bottomfraction}{1.0}
\section{Introduction}\label{Introduction}
Shortly before its perihelion passage on 16. Dec. 2011, Comet C/2011 W3 (Lovejoy) was detected by an Australian amateur astronomer. Quickly it turned out that this comet is a member of the Kreutz group, comets that approach the Sun within a few Solar radii. An extensive observational program was triggered involving all Solar observatories in space. Generally it was expected that Comet C/2011 W3 (Lovejoy) would not survive its perihelion passage at only $1.2 \, R_{sun}$ ($R_{sun}$: Solar radius). But Comet C/2011 W3 (Lovejoy) re-appeared from behind the Sun and disintegrated $1.6 \, \pm 0.2$ days after perihelion \citep{Sekanina2012}.
\par
The Kreutz group is a family of comets on similar highly inclined and high-eccentri\-city orbits, with semi-major axes of $\sim 100 \, \mathrm{AU}$ and periods of $\sim 1000 \, \mathrm{years}$. These comets are most probably the products of hierarchical breakups relating back to one big co\-metary nucleus with an estimated diameter of $\sim 100\,\mathrm{km}$ \citep{Kreutz1888, Sekanina2002a, Sekanina2004}.
\par
More than 2000 sun-grazing comets have been detected by the the Large Angle and Spectrometric Coronagraph (LASCO) onboard the Solar and Heliospheric Observatory (SOHO), most of them members of the Kreutz group. Almost daily a new comet is found by the sun pointing instruments on various Solar observatories \citep{Knight2010}. Generally the observations are not accurate enough to determine their orbital parameters, however, the inclination of the orbits are indicative for their membership.
\par
The vast majority of the sun-grazing group comets are small fragments (meter-sized to 10-meter-sized objects) that can only be detected very close to the Sun. Almost all are annihilated by the intense Solar radiation and particle bombardment within the corona, others hit the Sun directly. Recently, the demise of a comet within the lower corona could be observed directly for the first time \citep{Schri2012}. A few large comets of the Kreutz group were observed from ground. The spectacular Comet Ikeya-Seki (C/1965 S1) and the Great Comet of 1882 (C/1882 R1) are the most prominent ones. They probably have split from each other during a perihelion passage about 900 years ago \citep{Marsden1967}. Some comets arrive in pairs. Modeling of their dynamical behavior of the Kreutz group shows that these comets can split anywhere along their orbits \citep{Sekanina2007}. They seem to be distributed along the original orbit of their progenitor.
\par
We assessed the size of the nucleus of Comet C/2011 W3 (Lovejoy) following \citet{Knight2010}, who investigated the light curves of Kreutz group comets during their approach towards the Sun. The brightness of these comets peaks around a heliocentric distance of $10$ to $12 \, R_{sun}$. A comet with a radius of $4 \, \mathrm{m}$ shows a brightness of $8 \, \mathrm{mag}$ at $12 \, R_{sun}$. The peak brightness of Comet Lovejoy was estimated around $-4 \, \mathrm{mag}$ (Karl Battams, NRL, 2012)\footnote{$\mathrm{http:\slash\slash sungrazer.nrl.navy.mil\slash index.php?p=news\slash birthday\_comet\_c1}$}. This converts to a radius of $\sim 1 \, \mathrm{km}$ for the nucleus of Comet C/2011 W3 (Lovejoy). At the Comet Lovejoy workshop in Boulder Colorado (March 21-22, 2012), J. C. Raymond reported Lyman alpha observations of the hydrogen (water) production of comet Lovejoy using the UVSC spectrograph onboard the SOHO spacecraft \citep[for previous comet observations see e.g.][]{Ciaravella2010}. The derived estimate of the nucleus size was $0.4 \, \mathrm{km}$ in diameter shortly before perihelion.
\par
The few bright Kreutz group comets (observed from ground) are listed, e.g. in \citet{Knight2010}. Their orbital parameters can be found in the small bodies data base of JPL\footnote{$\mathrm{http: \slash \slash ssd.jpl.nasa.gov \slash sbdb.cgi}$}. All but one of these comets possess perihelia smaller than the Solar Roche limit for self-gravitating solid bodies,
\begin{equation}
r_{roche} \,  = \, R_{sun} \, \left( \, \frac{2 \, \pi}{3} \, \frac{\rho_{sun}}{\rho}   \,\right)^{1/3} \, = \, 1.8 \, R_{sun} \,\mathrm{,}
\label{eq_8}
\end{equation}
\citep[calculated for a density of the cometary nucleus of $\rho \, = \, 500 \, \mathrm{kg \, m^{-3}}$, which reflects a porous mixture of water ice and dust;][]{Davidsson2007, Thomas2009}. Here, $\rho_{sun}$ is the mean density of the Sun. The surprising fact that these comets survived their Solar passages requires additional forces beyond self-gravity to keep them together. Papers dealing with the fates of sun-grazing comets \citep{Huebner1967, Sekanina2003, Iseli2002, Brown2011} usually investigate processes that lead to the demise of the comets near the sun, but they do not investigate why some survive a passage deep inside the Roche limit. One obvious force is due to the tensile strength of the cometary material. Estimations of the tensile strength of cometary nuclei are ranging from high values, $\sim 1 \, \mathrm{kPa}$ \citep{Rickman1998,Yelle2004,Richardson2007}, to very low values, $\sim 1 \, \mathrm{Pa}$ \citep{SkorovBlum2011}. Recent observations and in particular the Deep Impact results \citep{Ahearn2005,Holsapple2007} have shown that the tensile strength of comets may be very low, clearly below $1 \, \mathrm{kPa}$, most probably in the range of $10 \, \mathrm{Pa}$ to $100 \, \mathrm{Pa}$. Model calculations of the breakup of Comet D/1993 F2 (Shoemaker-Levy), when it entered Jupiter's Roche limit, also supports the assumption of low cometary tensile strength \citep{Asphaug1994}. Such low values are, however, too small to explain the survival of the brightest Kreutz group comets.
\par
So why do big Kreutz group comets survive the passage within the Roche limit but not Comet  D/1993 F2 (Shoemaker-Levy)? One obvious difference is that Comet Lovejoy was extremely active near the Sun while Comet  D/1993 F2 (Shoemaker-Levy) was completely inactive at $5 \, \mathrm{AU}$ from the Sun. The reaction force caused by the strong outgassing (sublimation) of the nucleus near the Sun acts to keep the nucleus together and to overcome the tidal disruption. Due to the size dependence of this force (see Eq. \ref{eq_6}), it provides an upper limit for the nucleus size. A lower limit is set by the amount of eroded material during the perihelion passage. This limit has been widely discussed, e.g., by \citet{Huebner1967}, \citet{Weissman1983}, \citet{Iseli2002} and \citet{Sekanina2003}. The actual size of Comet C/2011 W3 (Lovejoy) probably was close to this lower limit. It survived the near Solar passage but its activity terminated in a cloud of dust at $0.4 \, \mathrm{AU}$ \citep[1.6 days after its perihelion,][]{Sekanina2012}.
\par
In this work, we will show a feasibility analysis that sungrazing comets can survive a close passage through the Solar corona due to the reaction forces of their high outgassing rate. Our model bears many simplifications, so that more detailed analysis of the assumptions and effects is necessary in a future study, but provides a sufficient criterion for stability, independent of the equation of state of cometary matter. Forces acting on sungrazing cometary nuclei are discussed in Sect. \ref{Forces acting on sungrazing comets within the Roche limit}. In Sect. \ref{Size estimation of surviving sungrazing comets within the Roche limit} an upper limit for the nucleus size is derived from the balance of forces within the Roche limit as a function of the heliocentric distance. Finally, the results are discussed in the concluding Sect. \ref{Conclusion}.

\section{Forces acting on sungrazing comets}\label{Forces acting on sungrazing comets within the Roche limit}
\begin{figure}[tb!]
\centering
\includegraphics[angle=0,width=0.7\columnwidth]{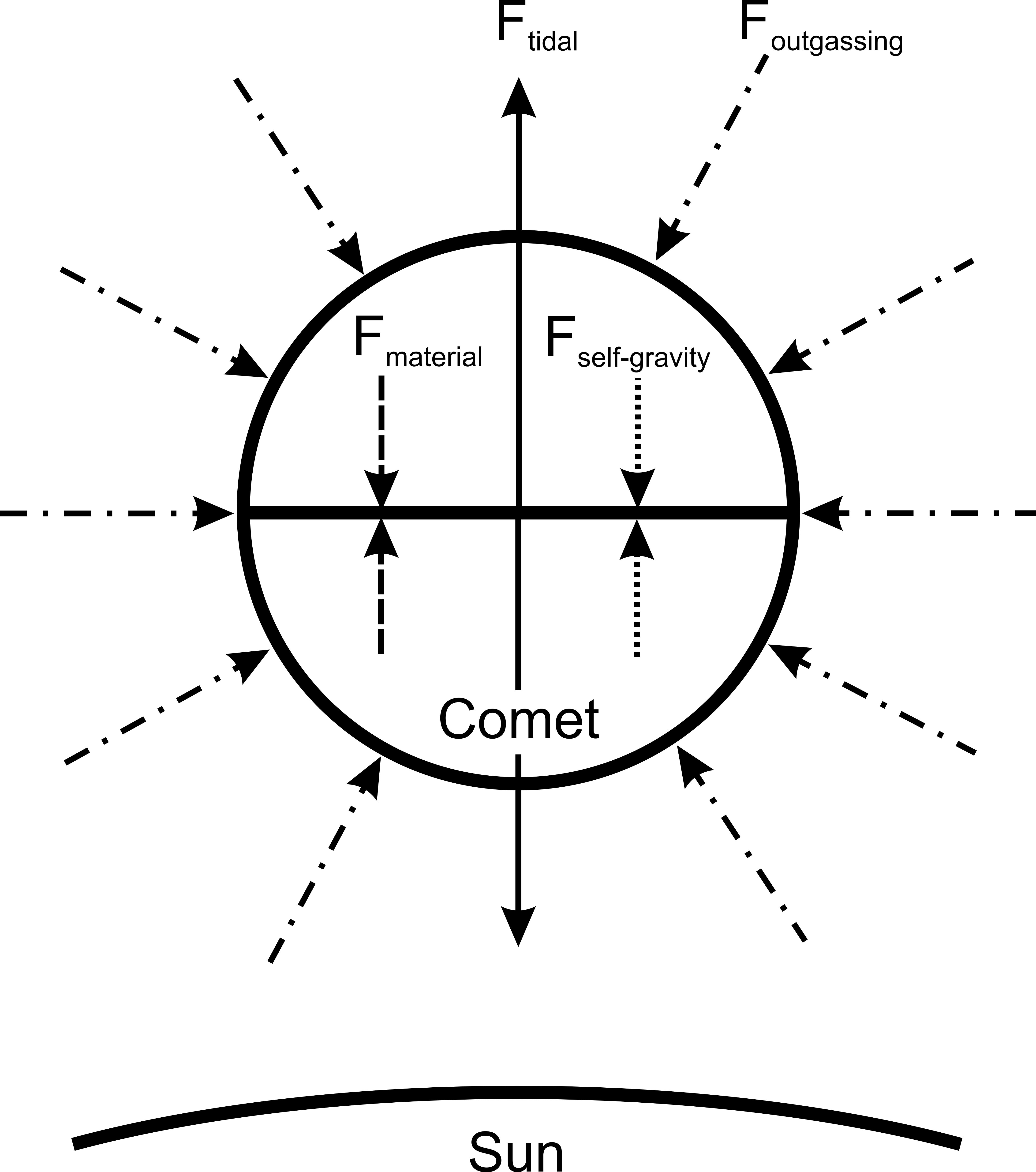}
\caption{Overview of the forces acting on sungrazing comets.}
\label{FigPaperI.0}
\end{figure}

In this section, relevant forces acting on sungrazing comets are discussed. Without better knowledge of Comet C/2011 W3 (Lovejoy) and for a more general use, we assume a spherical, isotropic cometary nucleus at perihelion. Further, we assume that all relevant forces can be referred to the midplane of the nucleus (see Fig. \ref{FigPaperI.0}). Our attempt is to show in a simple model that reaction forces due to the rapid outgassing close to the Sun's surface can considerably contribute to the cohesion and, thus, to the survival of comet nuclei at close encounters.
\par
Within the Roche limit, cometary nuclei can be disrupted due to tidal forces acting on the material. The tidal force acting on cometary nuclei was investigated in detail by \citet[][]{Davidsson2001}. Here, we adopted his formulation of the tidal force (with $\theta \, = \, 0$ and $f \, = \, 1$ for a spherical nucleus), i.e.
\begin{equation}
F_{tidal} \, =  \, \frac{1}{2} \, \frac{\pi \, \rho \, \gamma \, M_{sun} \, R^4}{r_P^3} \, \mathrm{,}
\label{eq1}
\end{equation}
where $M_{sun}$ is the solar mass, $\gamma$ is the gravitational constant, $\rho$ is the mean density of the nucleus, $R$ is the radius of the nucleus and $r_P$ is the heliocentric distance, or in our simplifying case the perihelion distance.
\par
The self-gravity of the cometary nucleus acts against the tidal forces caused by the gravity field of the Sun. Here, the expression derived by \citet[][with $A \, \approx \, 4$ and $f \, = \,1$ for a spherical nucleus]{Davidsson2001} is used to calculate the self-gravity of the cometary nucleus,
\begin{equation}
F_{self-gravity} \, =  \, - \,\pi \, \rho^2 \, \gamma \, R^4 \, \mathrm{,}
\label{eq1.1}
\end{equation}
here written as a negative quantity to indicate that this force is trying to prevent the tidal splitting.
\par
The material force further acts against the tidal splitting of the cometary nucleus. In our simple case, this force is given by the tensile strength of the material $p_{tensile}$ multiplied by the cross-sectional area of the nucleus at the midplane,
\begin{equation}
F_{material} \, = \, - \, \pi  \, p_{tensile} \, R^2 \, \mathrm{.}
\label{eq2}
\end{equation}
Observations and modeling of cometary nuclei yield a broad range of possible values for the tensile strength of the material. \citet{Richardson2007} estimated an tensile strength of $1 - 10 \, \mathrm{kPa}$ for the cometary material using results obtained by the Deep Impact mission. \citet{Holsapple2007} showed that tensile strengths from $0\, \mathrm{Pa}$ to $1 - 12 \, \mathrm{kPa}$ could fit the observational data. The tensile strength of the nucleus of Comet C/Hyakutake 1996 B2 must be at least $20 \, \mathrm{Pa}$ to $300 \, \mathrm{Pa}$ to avoid rotational breakup of the nucleus \citep{Lisse1999}. The breakup of Comet D/1993 F2 (Shoemaker-Levy) yield tensile strengths of $\sim 270\, \mathrm{Pa}$ \citep{Greenberg1995} and $\sim 1\, \mathrm{kPa}$ \citep{Rickman1998}. However, another model calculation of the breakup of Comet D/1993 F2 (Shoemaker-Levy) suggest vanishing low tensile strength of the material \citep{Asphaug1994}. Recently, theoretical estimates of the strength of the material composed of individual dust and ice aggregates were developed by \citet{SkorovBlum2011}. They found that the tensile strength of hierarchical porous materials can be much lower than that of homogeneous (but porous) dust-ice mixtures. For aggregate sizes of $1 \, \mathrm{mm}$, they find tensile-strength values of only a few Pa. Thus, reasonable values for the tensile strength of a microscopically inhomogeneous cometary nucleus composed of dust and water ice should fall in the range between $p_{tensile} \, = \,\sim 1 \, \mathrm{Pa}$ \citep[for $\sim  1 \, \mathrm{mm}$ sized aggregates and a volume filling factor of $\phi \, = \, 0.3$;][]{SkorovBlum2011} and $p_{tensile} \, = \, \sim 1 \, \mathrm{kPa}$ \citep[homogeneous dusty bodies at high porosity;][]{Blum2006a}.
\par
The sublimation of the icy constituents leads to an additional force acting against the separation of the nucleus. This force depends on the amount of absorbed energy of the cometary nucleus, given by the energy balance equation for the surface layers of the nucleus. We assume that the dominant volatile on the nucleus surface is water ice, and that the surface is uniform and is not covered with a substantial dust crust. For high surface temperatures \citep[$T \, \gtrsim 180 \, \mathrm{K}$; ][]{Huebner1967,Sekanina2012} the absorbed energy is essentially consumed by the sublimation process. The heat conduction into the interior of the nucleus and the thermal emission are much smaller. In this case, the energy balance equation for the surface of the nucleus is
\begin{equation}
\frac{1}{4} \,\xi \, E_{sun}(r_P) \, = \, Z(T) \, \Lambda(T)  \, \mathrm{.}
\label{eq3}
\end{equation}
The factor $1 \, / \, 4$ arises from the assumption that the passing radiative flux is uniformly redistributed to the entire surface by the coma around the nucleus. This assumption is motivatied by the fact that a coma around a cometary nucleus can significantly influence the amount and the distribution of the incident radiation \citep{Hellmich1981}. Here,
\begin{equation}
E_{sun}(r_P) \, = \, \pi \,F  \, \left( \, \frac{R_{sun}}{r_P} \, \right)^2 \, \frac{\chi_1}{\chi_2}  \, \mathrm{,}
\label{eq3.1}
\end{equation}
is the energy flux density from the Sun at the position of the cometary nucleus, with $\pi \, F \, = \, 6.37 \times 10^7 \, \mathrm{W \, m^{-2}}$, $\chi_1 \, = \, ( \, 1 \, - \, [ \, 1 \, - \, (\, R_{sun}^2 \, / \ r_p^2 \, ) \, ]^{1/2} \, ) \, / \, 2$ and $\chi_2\, = \, R_{sun}^2 \, / ( \, 4  \, r_P^2\,)$ \citep{Huebner2007}. This formulation takes the finite energy flux close to the Sun due to the extended solar surface into account. $Z(T)$ and $\Lambda(T)$ are the sublimation rate and the latent heat of water ice, respectively. The factor $\xi$ was added to the model to vary the absorbed energy by the surface layers of the nucleus. The variation of the absorbed energy can have different causes:
\begin{enumerate}
\item The albedo of the surface materials influences the amount of absorbed energy by the surface layers.
\item The presence of an optically thick coma can lead to an increase of the effective absorbing cross section of the comet \citep{Hellmich1981}. However, the energy distribution inside the coma can either increase, or decrease the radiative flux to the surface of the nucleus.
 \item Additional energy input can also be caused by the solar wind, hitting the cometary nucleus \citep{Brown2011}.
\end{enumerate}
\par
For water ice, the sublimation rate can be calculated using the Hertz-Knudsen formula \citep{Knudsen1909},
\begin{equation}
Z(T) \, = \, \alpha(T) \, p_{sat}(T) \, \sqrt{\frac{m}{2 \, \pi \, k \, T}}  \, \mathrm{,}
\label{eq4}
\end{equation}
where $\alpha(T)$ is the sublimation coefficient, $m$ is the mass of a water molecule, $k$ is the Boltzmann's constant and $p_{sat}(T)$ is the saturation pressure of water ice \citep[see][and references therein for details]{Gundlach2011}.
\par
The resulting force caused by the outgassing of the water ice stems from the momentum transfer of the water molecules leaving the surface and can be written as the product of the gas pressure and the cross-sectional area of the nucleus at the midplane,
\begin{equation}
F_{outgassing} \, = \, - \, \pi  \, p_{sat}(T) \, R^2 \, \mathrm{.}
\label{eq_6}
\end{equation}
The temperature is calculated from the energy balance given by Eq. \ref{eq3}.
\par
Simple estimates show that the centrifugal force can be neglected for the slow rotator and for typical sizes of cometary nuclei.

\section{Size estimation of surviving sungrazing comets within the Roche limit}\label{Size estimation of surviving sungrazing comets within the Roche limit}
With estimates of all forces in place (Eqs. \ref{eq1}, \ref{eq1.1}, \ref{eq2}, \ref{eq_6}), the reaction of the comet can be estimated both dynamically and statically. A dynamical requirement for survival of the comet nucleus can be written in the form
\begin{equation}
F_{tidal} \, + \, F_{self-gravity} \, + \, F_{material} \, + \, F_{outgassing} \, \leq \, M_{comet} \, v_{esc} \, c \, \tau^{-1} \, \mathrm{,}
\label{eq_6.1}
\end{equation}
with $M_{comet}$, $v_{esc}$, $c$, and $\tau$ being the comet mass, the escape speed from the comet's surface, a factor $\sim 1$, and the time of perihelion passage, respectively. If Eq. \ref{eq_6.1} is satisfied, then the comet material is unable to leave the comet's Hill sphere during the time of passage. A stronger requirement for stability, however, is the condition that the comet is even statically stable. This is the case if
\begin{equation}
F_{tidal} \, + \, F_{self-gravity} \, + \, F_{material} \, + \, F_{outgassing} \, \leq \, 0 \, \mathrm{.}
\label{eq_7}
\end{equation}
Here, the cometary material never leaves the comet nucleus surface because the compressing gravitational, material and outgassing forces exceed the tidal force at any time. This criterion is independent of the dynamical constitutive laws of the comet material (unlike Eq. \ref{eq_6.1}) so that it can be applied to comets (whose material properties and constitutive laws are basically unknown) with only the tensile strength as a free parameter.
\par
Eq. \ref{eq_7} can be used to derive the maximum radius of a cometary nucleus able to sustain the tidal splitting within the Roche limit in the static case. Even for low tensile strength of the cometary nucleus, tidal breakup can be prevented by the reaction force caused by the sublimation of the icy constituents. Dynamical separation of possible fragments due to dynamical evolution is not considered in this approach.
\par
\begin{figure}[tb!]
\centering
\includegraphics[angle=180,width=1.0\columnwidth]{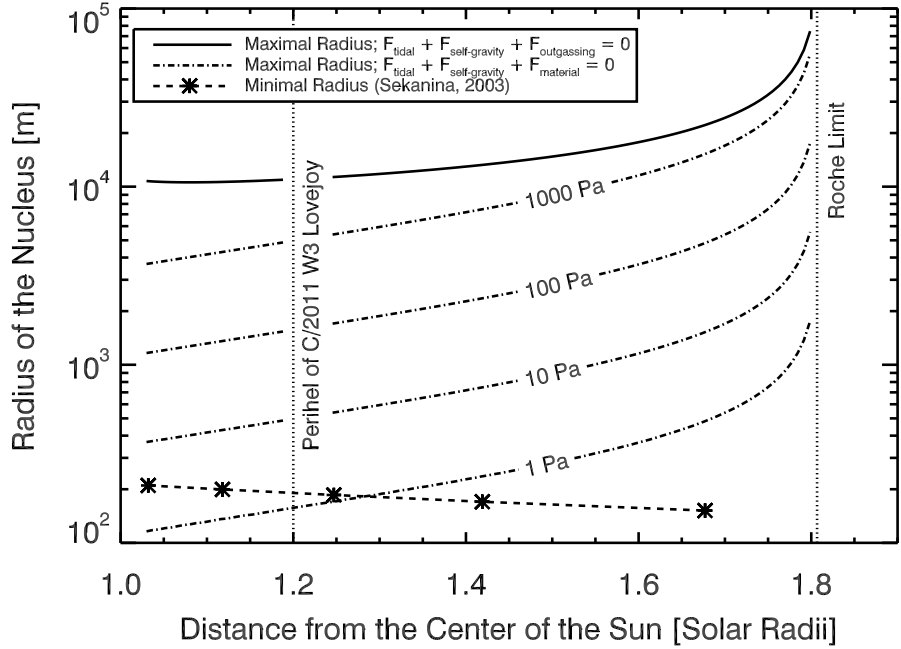}
\caption{Maximum radius of a cometary nucleus able to survive within the Roche limit as a function of the perihelion distance. The solid curve shows the resulting maximum radius using Eq. \ref{eq_7} (with $\xi \, = \, 1$), but neglecting the material force. Resulting maximum radii using Eq. \ref{eq_7} and different values for the tensile strength,  $p_{tensile} \, = \, 1 \, \mathrm{Pa}$, $p_{tensile} \, = \, 10 \, \mathrm{Pa}$, $p_{tensile} \, = \, 100 \, \mathrm{Pa}$, $p_{tensile} \, = \, 1000 \, \mathrm{Pa}$, but neglecting the reaction force caused by the outgassing, are shown by the dasched-dotted curves. For comparison, the lower limit on the radius is given by the thickness of the eroded layer during the preperihelion passage and the postperihelion branch of the orbit \citep[asterisks and dashed curve;][]{Sekanina2003}. The dotted vertical lines are denoting the perihelion distance of Comet C/2011 W3 (Lovejoy) and the Roche limit for a density of $\rho \, = \, 500 \, \mathrm{kg \, m^{-3}}$.}
\label{FigPaperI.1}
\end{figure}

Fig. \ref{FigPaperI.1} shows the resulting maximum radius of the nucleus for different perihelion distances (solid curve), using the force caused by the outgassing (Eq. \ref{eq_6} with $\xi \, = \, 1$) together with the self-gravity force (Eq. \ref{eq1.1}), but neglecting material strength (i.e. $F_{material} \, = \, 0$). The resulting maximum radius of a cometary nucleus, to survive within the Roche limit, due to the material force (Eq. \ref{eq2}) and the self-gravity force (Eq. \ref{eq1.1}), without the force caused by the outgassing is shown by the dashed-dotted curves, for $p_{tensile} \, = \, 1 \, \mathrm{Pa}$, $p_{tensile} \, = \, 10 \, \mathrm{Pa}$, $p_{tensile} \, = \, 100 \, \mathrm{Pa}$, $p_{tensile} \, = \, 1000 \, \mathrm{Pa}$ (see Sect. \ref{Forces acting on sungrazing comets within the Roche limit}). The mean density of the nucleus was assumed to be $\rho \, = \, 500 \, \mathrm{kg \, m^{-3}}$, which reflects a porous mixture of water ice and dust \citep{Davidsson2007, Thomas2009}.
\par
It is shown in Fig. \ref{FigPaperI.1} that the force caused by the outgassing can be orders magnitude higher than the material force. Thus, active comets are able to withstand tidal disruption, even if the tensile strength of the material is negligibly low.
\par
The lower limit on the radius of the nucleus is given by the thickness of the eroded layer during the preperihelion passage and the postperihelion branch of the orbit. The thickness of the eroded layer of sungrazing comets was estimated e.g. by \citet{Sekanina2003} for different perihelion distances and a mean density of the nucleus of $\rho \, = \, 150 \, \mathrm{kg \, m^{-3}}$. For comparison, these results were recalculated to a mean density of $\rho \, = \, 500 \, \mathrm{kg \, m^{-3}}$ (shown by the asterisks and the dashed curve in Fig. \ref{FigPaperI.1}). The perihelion distance of Comet C/2011 W3 (Lovejoy) and the Roche limit for a density of $\rho \, = \, 500 \, \mathrm{kg \, m^{-3}}$ are denoted by the dotted vertical lines.
\par
Using Eq. \ref{eq_7} for Comet C/2011 W3 (Lovejoy; perihelion distance: $r_{p,LJ} \, = \, 1.2 \, R_{sun}$) yields a maximal radius of $R_{max,LJ} \, = \, 11.0 \, \mathrm{km}$ (for $\xi \, = \, 1$ and $p_{tensile} \, = \, 0 \, \mathrm{Pa}$). The minimal radius of Comet C/2011 W3 (Lovejoy) is given by $R_{min,LJ} \, = \, 0.2 \, \mathrm{km}$. A variation of the absorbed energy from $\xi \, = \, 0.5$ ($50 \, \%$ less energy absorbed by the surface layers of the nucleus) to $\xi \, = \, 1.5$ ($50 \, \%$ more energy absorbed by the surface layers of the nucleus) leads to maximum radius of Comet C/2011 W3 (Lovejoy) of $R_{max,LJ} \, = \, 7.7 \, \mathrm{km}$ and $R_{max,LJ} \, = \, 13.6 \, \mathrm{km}$, respectively.

\section{Discussion and conclusion}\label{Conclusion}
In this work, we have shown that sungrazing comets are able to survive within the Roche limit due the reaction force caused by the outgassing of the icy constituents, even if the tensile strength of the material is low ($p_{tensile} \,< \,  1 \, \mathrm{kPa}$). The assumption that the entire surface of the nucleus is composed of pure water ice yields an upper limit for the radius of the nucleus able to survive within the Roche limit of the Sun, due to the self-gravity force and the reaction force caused by the outgassing.
\par
Using this approach, we derive a maximum radius of $R_{max,LJ} \, = \, 11.0\, \mathrm{km}$ (for $\xi \, = \, 1$, $p_{tensile} \, = \, 0 \, \mathrm{Pa}$ and $\rho \, = \, 500 \, \mathrm{kg \, m^{-3}}$; see Fig. \ref{FigPaperI.1}) for Comet C/2011 W3 (Lovejoy). The variation of the energy available for the sublimation process from $\xi \, = \, 0.5$ to $\xi \, = \, 1.5$ (see Eq. \ref{eq3}) yields $R_{max,LJ} \, = \, 7.7 \, \mathrm{km}$ and $R_{max,LJ} \, = \, 13.6\, \mathrm{km}$, respectively.
\par
The lower limit on the radius of the nucleus is given by the thickness of the eroded layer during the preperihelion passage and the postperihelion branch of the orbit. Thus, the nucleus of Comet C/2011 W3 (Lovejoy) must have been bigger than $R_{min,LJ} \, = \, 0.2 \, \mathrm{km}$ in order to provide enough material for the erosion. The radius estimated from the observed brightness ($R_{LJ} \, \approx \, 1 \, \mathrm{km}$, see Sect. \ref{Introduction}) and the radius derived from Lyman alpha observations ($R_{LJ} \, = \, 0.2 \, \mathrm{km}$, see Sect. \ref{Introduction}), are in qualitative agreement with our results.
\par
Due to the outgassing of the icy constituents, the maximum radius of sungrazing comets able to survive within the Roche limit is relatively large (see Fig. \ref{FigPaperI.1}). However, if the effective gas production decreases, the outgassing force decreases and, therewith, the maximum radius of the nucleus able to survive within the Roche limit also decreases. Thus, cometary nuclei with low tensile strength can only survive within the Roche limit if they are active. Members of the Kreutz group comets like Comet C/2011 W3 (Lovejoy) are probably young fragments of a big progenitor comet \citep{Sekanina2002c} and are, thus, active.
\par
Two very big Kreutz group comets, 1882 II and 1963 V were observed within the Roche limit of the Sun \citep[$R_{1882II} \, = \, \sim 31\, \mathrm{km}$ and $R_{1963V} \, = \, \sim 14 \, \mathrm{km}$; ][]{Knight2010}. The perihelion distances were $r_{p,1882II} \, = \, 1.67 \, R_{sun}$ and $r_{p,1963V} \, = \, 1.09 \, R_{sun}$, respectively. Comet 1882 II had broken into at least five fragments \citep{Gill1883}. This observation is in agreement with our model, because the estimated radius of Comet 1882 II was bigger than the derived maximum radius for the survival of sungrazing comets within the Roche limit (see Fig. \ref{FigPaperI.1}). The survival of comet 1963 V can be explained with our model within the error of our model and the error of the size estimation.
\par
A considerably higher activity \citep[caused by, e.g., volatiles with a much higher saturation pressure; e.g. $~ 20 \%$ of the global gas production rate of comet 109P/Hartley was $\mathrm{CO_2}$;][]{AHearn2011}, or additional forces, e. g. the reaction force caused by the momentum transfer of the solar wind, can additionally increase the maximum radius of sungrazing comets able to withstand tidal disruption within the Roche limit.
\par
In the case of hemispheric activity, half of the surface receives the total amount of the incoming energy. This leads to an increased sublimation pressure and, thus, to an increased reaction force caused by the outgassing. However, this resulting force only acts on one hemisphere. Thus, the assumption of hemispheric activity does not qualitatively change the occurrence of this effect, but can cause a slightly different maximum radius of the nucleus.
\par
Unfortunately, the survival of sungrazing comets within the Roche limit does not provide any information about the tensile strength of the cometary nucleus as long as the the reaction force caused by the outgassing exceeds the material force.

\bibliographystyle{model2-names}
\bibliography{bib}

\end{document}